\documentclass[aps,pra,twocolumn,showpacs,floatfix]{revtex4}
\usepackage{amssymb}    
\usepackage{graphicx}   
\usepackage{epsfig}
\usepackage{verbatim}   
\usepackage{color}      
\usepackage{subfigure}  
\usepackage{hyperref}   

\newcommand{\CO}{(color online)\hspace{1mm}}

\begin{document}

\title{Ground state phase diagram of the repulsive SU(3) Hubbard model in Gutzwiller approximation}
\author{\'Akos  Rapp}\affiliation{ Institut f\"ur Theoretische Physik, Universit\"at zu K\"oln, D-50937 K\"oln, Germany}
\author{Achim Rosch}\affiliation{ Institut f\"ur Theoretische Physik, Universit\"at zu K\"oln, D-50937 K\"oln, Germany}

\date{\today}

\begin{abstract}
We perform a variational Gutzwiller calculation to study the ground state of the repulsive SU(3) Hubbard model on the Bethe lattice with infinite coordination number. We construct a ground-state phase diagram  focusing on phases with a two-sublattice structure and find five relevant phases: (1) a paramagnet, (2) a completely polarized ferromagnet, (3) a two-component antiferromagnet where the third component is depleted, (4) a two-component antiferromagnet with a metallic third component (an ``orbital selective'' Mott insulator), and (5) a density-wave state where two components occupy dominantly one sublattice and the last component the other one. First-order transitions between these phases lead to phase separation.  A comparison of the 
SU(3) Hubbard model to the better-known SU(2) model shows that the effects of doping are completely different in the two cases.
\end{abstract}


\maketitle

\section{Introduction}

The recent success in loading an ultracold fermionic cloud in an optical lattice with a total of eight fermion components and SU(2) $\times$ SU(6) global symmetry using alkaline-earth-metal atoms\cite{alkalineearth} demonstrates that fermionic SU($N$) Hamiltonians can be realized experimentally. A simple way to create SU($N$) symmetric models is to use the fact that neither the kinetic nor the interaction energy of atoms with closed electronic shells depend on the orientation of the nuclear spin.
For a given nuclear spin $I$ one can therefore realize SU($N$)-symmetric models with $N\leq 2 I+1$ by selectively occupying $N$ of the $2 I+1$ spin states \cite{2orbit-SUN-NPhys}.
For deep enough optical lattices the system is naturally described by a Hubbard model with a good accuracy\cite{Jaksch,BlochZwergerreview}. Generalizations of the Heisenberg and Kondo Hamiltonians have also been proposed \cite{2orbit-SUN-NPhys} to describe the system in other parameter regimes. While the $N=2$ Hubbard model has been studied extensively (see \cite{Hubbardreview} and references therein) and many results were obtained in large-$N$ expansions \cite{largeNlimit} (mainly focusing on even values of $N$ and the vicinity of half filling), much less is known for the SU($N$) case (far) away from half filling and if $N$ is odd. Experimentally, the first indications of the Mott transition in the SU(2) Hubbard model have recently been observed using fermionic atoms in optical lattices \cite{EsslingerHubbard,BlochHubbard} but lower temperatures are needed to stabilize, for example, an antiferromagnetic phase. 

In this work we shall focus on the ground state of $N=3$ fermionic components (colors) in a lattice with isotropic, local interactions. The case of attractive interactions have already drawn considerable attention \cite{Rapp,su3attr1,su3attr2,su3attr3} due to similarities to quarks \cite{Wilczek}, however, many-body losses (e.g., \cite{Li6exp} for $^6$Li) and high temperatures make experimental observation of the corresponding color superfluid phase and the liquid of the three-body bound states (``baryonic'' phase) difficult. In the case of the repulsive interactions, Szirmai, Legeza, and S\'olyom \cite{SUNDMRG} used DMRG and bosonization to study the phase diagram in one dimension, $D=1$. Already some time ago Honerkamp and Hofstetter \cite{HonerkampHofstetter} used functional renormalization group (fRG) in the $D=2$ square lattice close to half filling to investigate the interplay of spin- and charge-density waves and the possibility of staggered flux phases for larger $N$. More recently, Gorelik and Bl\"umer \cite{GorelikBlumer} performed paramagnetic dynamical mean field theory (DMFT) calculations to discuss the Mott phases. Finally, Miyatake, Inaba, and Suga \cite{su3half} studied two-sublattice ordering and Mott transitions at half filling as a function of the anisotropy in the interactions between the components using DMFT. 
Nevertheless, a comprehensive study of the phase diagram for arbitrary fillings is still missing. We apply a non-perturbative method, Gutzwiller ansatz with Gutzwiller approximation, to investigate the $T=0$ phase diagram which could serve as a starting point for future study. The Gutzwiller approach is a method which captures the physics both at weak and strong coupling and has strongly influenced our understanding of correlated matter \cite{Gutzwiller,SU2BrinkmanRice,MetznerVollhardt,SU2HubbardGutzwiller}. We concentrate on phases with two-sublattice ordering.

We will approximate the ground state of the Hamiltonian
\begin{equation}
	\hat H = -\frac{t^*}{\sqrt{z}}\!\!\sum_{\langle ij \rangle;\alpha}\!\!\left( \hat c_{i\alpha}^\dagger \hat c_{j\alpha} + {\rm h.c.} \right) + \frac{U}{2}\!\!\sum_{i;\alpha \neq \beta }\!\!\hat n_{i\alpha} \hat n_{i\beta} \;, \label{eq:H_0}
\end{equation}
where $\alpha=1,\;2,\;3$ denotes the three fermionic components, $i$ denotes sites in a bipartite lattice with $z$ nearest neighbors, $t^*$ is a normalized nearest-neighbor hopping amplitude, and $U > 0$ is a local repulsive interaction. We shall calculate ground state properties as a function of $U$ and the filling $\rho$, where $\rho=1$ corresponds to total filling, that is, three fermions per site.

The Hamiltonian in Eq.~(\ref{eq:H_0}) has a global SU(3) symmetry due to the isotropic interaction strengths between the components. Our main goal is to study how this symmetry breaks in the ground state due to the competition between kinetic and interaction energy. The scaling of the hopping ensures that the kinetic energy of Eq.~(\ref{eq:H_0}) per lattice site becomes independent of $z$ in the limit $z \to \infty$. As we will see, the Gutzwiller approximation, used below, is exact in this limit \cite{MetznerVollhardt}.

The Hamiltonian (\ref{eq:H_0}) can easily be generalized to SU($N$) global symmetry using $N$ components instead of three. We will mainly study the $N=3$ case, but consider arbitrary $N$ in some cases.

It is easy to see that the model has a particle-hole symmetry as the canonical transformation $c_{i\alpha} \to (-1)^i c_{i\alpha}^\dagger$ leaves the Hamiltonian invariant up to a constant term. Thus the phase diagrams for $\rho \leq 1/2$ and $\rho \geq 1/2$ can be mapped onto each other.

\section{Gutzwiller ansatz}

We approximate the ground state of the Hamiltonian in Eq.~(\ref{eq:H_0}) by a Gutzwiller wavefunction,
\begin{equation}
	\vert G \rangle = \prod_i \left( \sum_I \lambda_I(i) \hat p_{i I} \right) \; \vert \Psi_0 \rangle . \label{eq:G_0}
\end{equation}

The unprojected wavefunction $\vert \Psi_0 \rangle$ is a normalized Slater determinant. For simplicity, we assume that 
\begin{equation}
	\langle \Psi_0 \vert \hat c_{i\alpha}^+ \hat c_{i\beta} \vert \Psi_0 \rangle = \delta_{\alpha\beta} \langle  \Psi_0 \vert \hat n_{i\alpha} \vert \Psi_0 \rangle ,
\end{equation}
that is, we consider only collinear order where the magnetization matrices $\langle \Psi_0 \vert \hat c_{i\alpha}^+ \hat c_{i\beta} \vert \Psi_0 \rangle$ on different sites commute.

We shall use the notation
\begin{equation}
	n^0_{i\alpha} \equiv \langle \Psi_0 \vert \hat n_{i\alpha} \vert \Psi_0 \rangle .
\end{equation}

The operator $\hat p_{I}$ projects on the local configuration $I\in\left\{\emptyset,1,2,3,\bar 3 \equiv 12,\bar 2 \equiv 13, \bar 1 \equiv 23, t \equiv 123 \right\}$, satisfying $\hat p_I \hat p_{I'} = \delta_{II'} \hat p_I$. The Gutzwiller parameter $\lambda_I(i)$ therefore changes the relative amplitude of the configuration $I$ at site $i$. It is useful to define \emph{occupation} operators $\hat n_I$, with $\hat n_I \hat n_{I'} = \hat n_{I \cup I'}$. The two operator sets are related by
\begin{equation}
	 \hat n_{I} = \sum_{I' \supseteq I} \hat p_{I'} \;; \hat p_I = \sum_{I' \supseteq I} (-1)^{\vert I' \vert - \vert I \vert} \hat n_{I'} . \label{eq:def:p-n}
\end{equation}

The Gutzwiller expectation value of an operator $\hat O$ is defined as
\begin{equation}
	O \equiv \frac{\langle G \vert \hat O \vert G \rangle}{\langle G \vert  G \rangle}  . \label{eq:def:OG}
\end{equation}

We shall calculate the variational energy, defined by
\begin{equation}
	E_v \equiv H ,\label{eq:def:Evar}
\end{equation}
and minimize it with respect to $\vert \Psi_0 \rangle$ and $ \lambda_I(i)$.

\section{Gutzwiller approximation}

Unfortunately, the exact analytic evaluation of the Gutz\-wil\-ler expectation values in Eq.~(\ref{eq:def:Evar}) is in general not possible. Originally, Gutzwiller \cite{Gutzwiller} proposed a mean-field-like approximation to calculate the variational energy. Later it was shown in the $N=2$ case that in the limit of infinite coordination numbers, $z\to \infty$, the Gutzwiller approximation is exact \cite{MetznerVollhardt,Gebhard}. Here we shall outline an approach to calculate the variational energy in Gutzwiller approximation in the $N=3$ case based on functional integrals. Technical details can be found in the Appendix. An alternative method to derive the variational energy within the Gutzwiller approximation has been described for multiband models by B\"unemann, Gebhard, and Weber in Ref.~\cite{BunemannGebhardWeber}.

The main idea is to express all Gutzwiller expectation values as expectation values in a \emph{static} auxiliary field theory, as in Ref.~\cite{Rapp}. The simplest way to construct this field theory is to start from the norm of the Gutzwiller wave function. We observe that after normal ordering and then using Wick's theorem we can always write~\cite{Rapp}
\begin{eqnarray}
	\langle G \vert G \rangle &=& \int {\cal D}\Psi^\dagger {\cal D}\Psi \; e^{\Psi^\dagger G_0^{-1} \Psi} \nonumber \\
	&& \times \prod_i \left[ \sum_I \left( \sum_{I' \subseteq I} (-1)^{\vert I \vert - \vert I' \vert} \lambda_{I'}^2(i) \right) \tilde n_{i I} \right]\!\!, \label{eq:preSaux}
\end{eqnarray}
where $[G_{0}]_{i\alpha;j\beta} = \langle \Psi_0 \vert \hat c_{j\beta}^\dagger \hat c_{i\alpha} \vert \Psi_0 \rangle$ is the bare equal-time propagator of the unprojected wave function, $[\Psi]_{i\alpha} = c_{i\alpha}$ is a time-independent Grassmann field, and the occupations, $\tilde n_{i I} = \prod_{\alpha \in I} \bar c_{i \alpha} c_{i \alpha}$, are expressed  in terms of the Grassmann variables $c_{i \alpha}$. Using the Grassman algebra, we can exponentiate the terms in the product, and rewrite Eq.~(\ref{eq:preSaux}) as a ``partition function'',
\begin{eqnarray}
	\langle G \vert G \rangle = Z_{\rm aux} &=& \int {\cal D}\Psi^\dagger {\cal D}\Psi \; e^{-S_{\rm aux}[\Psi^\dagger,\Psi]} , \nonumber \\
	S_{\rm aux}[\Psi^\dagger,\Psi] &=& \Psi^\dagger (- G_0^{-1}) \Psi + \sum_{i;I} u_I(i) \tilde n_{i I} ,\label{eq:def:Saux}
\end{eqnarray}
where $u_I(i)$ are functions of the $\lambda_I(i)$. It can be seen that for any operator, $\langle  G \vert \hat O \vert G \rangle$ can be expressed as some functional integral with the same static action $S_{\rm aux}$. As a consequence, Gutzwiller expectation values become certain expectation values in an interacting field theory. We can calculate all such expectation values in the limit of $z \to \infty$ using the cavity method \cite{DMFT} described in the Appendix. Here we use that to obtain the Gutzwiller approximation for the {\it static} field theory $S_{\rm aux}$, one follows the same steps which have to be taken to derive
dynamical mean field theory (DMFT) \cite{DMFT} in the large $z$ limit.
 As there is no time dependence, the relevant Grassmann integrations become finite dimensional and can be calculated analytically (see the Appendix).
Note however, that the coupling parameters of $S_{\rm aux}$  remain variational parameters and one still has to minimize the resulting energy.

In the limit of $z\to \infty$, the proper self-energy matrix $\Sigma = G_0^{-1} - \langle - \Psi \Psi^\dagger \rangle_{S_{\rm aux}}^{-1}$ corresponding to the action $S_{\rm aux}$ becomes site diagonal $\Sigma_{ij \alpha} = \delta_{ij} \Sigma_{i \alpha}$ as the arguments of Ref.~\cite{MetznerVollhardt} can be generalized to include higher order local interaction terms. 
It is important to note that in contrast to DMFT, the self energy is \textit{static}, that is, it does not depend on frequency. Therefore all self-energy effects
can be absorbed into $G_0$. Thus, by restricting the variational parameters $\lambda_\emptyset(i)$ and $\lambda_\alpha(i)$ such that
\begin{eqnarray}
	\langle G \vert G \rangle &=& 1 \;{\rm and} \label{eq:cond-1} \\
	n_{i\alpha} &=& \langle \Psi_0 \vert \hat n_{i\alpha} \vert \Psi_0 \rangle = n_{i\alpha}^0 \; (\forall \alpha), \label{eq:cond-2}
\end{eqnarray}
one can set the self energy to zero ($\Sigma_{i\alpha} = 0$). 

Note that in Eq.~(\ref{eq:cond-2}) we simply require that the Gutz\-wil\-ler projection has to leave the densities invariant. As a consequence, the parameters $\lambda_\alpha(i)$, $\alpha=1,\;2,\;3$, are simply replaced by the order parameters of $\vert \Psi_0 \rangle$ as variational parameters. Furthermore, it is also possible to replace the two- and three-body Gutz\-wil\-ler parameters, $\lambda_{\bar \alpha}(i)$ and $\lambda_t(i)$, by the physical occupation numbers, $d_{i\alpha} \equiv n_{i \bar \alpha}$ and $t_i \equiv n_{i t}$ as variational parameters (see the Appendix for details).

After some algebra, the variational energy can be cast to the form
\begin{eqnarray}
	E_v &=& -\frac{t^*}{\sqrt{z}}\!\!\sum_{\langle ij \rangle;\alpha}\!\!q_{i\alpha} q_{j\alpha} \langle \Psi_0 \vert \hat c_{i\alpha}^\dagger \hat c_{j\alpha} \vert \Psi_0 \rangle + {\rm H.c.}  \nonumber \\
	&& \qquad \qquad \qquad \qquad \qquad  + \; U \sum_{i;\alpha} d_{i\alpha} , \label{eq:Evar-gen}
\end{eqnarray}
where the renormalization factors are given by (shown here for $\alpha = 1$, other components are related by permutations)
\begin{eqnarray}
	q_{i\,1} &=& \frac{ \sqrt{p_{i \emptyset} p_{i 1} } + \sqrt{p_{i 2} p_{i \bar 3}} + \sqrt{p_{i 3} p_{i \bar 2} } + \sqrt{p_{i \bar 1} p_{i t}}  }{ \sqrt{ n_{i 1}^0 (1 - n_{i 1}^0 ) } } . \label{eq:def:qi1}
\end{eqnarray}
Here, $p_{i I}= \langle G \vert \hat p_{i I} \vert G \rangle/\langle G \vert  G \rangle $ are functions of the physical occupancies $n_{i I}$ [see Eq.~(\ref{eq:def:p-n})]. Note that for any physical solution, $q_{i \alpha} $ has to be real, and thus there is a number of constraints on the variational space corresponding to 
\begin{equation}
p_{i I} \geq 0 . \label{eq:constraints}	
\end{equation}

Equation (\ref{eq:Evar-gen}) is consistent with the general form of the variational energy in Gutzwiller approximation obtained by using other techniques (see Refs.~\cite{BunemannGebhardWeber,RadnotiGebhardFazekas}). The physical interpretation of the renormalization factors $q_{i\alpha}$ is that they describe how the noninteracting band gets renormalized by the interactions. 

Finally, for a fixed set of the bare occupations $\{ n^0_{i\alpha} \}$, the variational problem can be written in the form
\begin{eqnarray}
	{\rm min} && \Big[ E_v - E_{\rm sp} \big(\langle \Psi_0 | \Psi_0 \rangle - 1\big) \nonumber \\
		&& \quad +\sum_{i\alpha} \lambda_{i\alpha} \big( n^0_{i\alpha} - \langle \Psi_0 | \hat n_{i\alpha} | \Psi_0 \rangle \big) \Big]  .\label{eq:Evar-G0}
\end{eqnarray}
Here $ E_{\rm sp}$ and $\lambda_{i\alpha}$ are Lagrange multipliers and $E_v$ is a function of $| \Psi_0 \rangle$ and the double and triple occupancies, $d_{i\alpha},t_i$. We observe that this expression is \emph{quadratic} in the unprojected wave function $| \Psi_0 \rangle$. As in Ref.~\cite{RadnotiGebhardFazekas}, one can perform the variation with respect to  $\langle \Psi_0 \vert$ analytically. This leads to an effective mean-field Schr\"odinger equation 
\begin{eqnarray}
\tilde H_0 | \Psi_0 \rangle &=& E_{\rm sp} | \Psi_0 \rangle , \\
\tilde H_0 &=& -\frac{t^*}{\sqrt{z}}\!\!\sum_{\langle ij \rangle;\alpha} \!\!q_{i\alpha} q_{j\alpha} \hat c_{i\alpha}^\dagger \hat c_{j\alpha}+  {\rm H.c.} - \sum_{i\alpha} \lambda_{i\alpha} \hat n_{i\alpha}, \nonumber
\end{eqnarray}
with $| \Psi_0 \rangle$ being the ground state of $\tilde H_0$. Therefore the unprojected wave function $| \Psi_0 \rangle$ is a Slater determinant as required by our approach. For a specific set of  $\{ n^0_{i\alpha} \}$ (these also define in our case the order parameters in the unprojected state), one could in principle construct the unprojected wave function. However, we do not need $|\Psi_0\rangle$ explicitly, only the corresponding ground-state expectation values. For this reason, we shall perform calculations on the Bethe lattice with infinite coordination number, where the (ground state) Green's function corresponding to $\tilde H_0$ can be calculated analytically. Another advantage of doing so is that on the homogeneous Bethe lattice, the density of states (DOS) is semielliptic. This, on one hand, is a good approximation to the density of states of a $d=3$ simple cubic lattice with nearest-neighbor hopping, while on the other hand, it allows us to compare our results quantitatively to DMFT studies, which used the same bare DOS (see Refs.~\cite{GorelikBlumer} and \cite{su3half}).

\section{Two-sublattice order on the Bethe lattice}

The variational problem in Eq.~(\ref{eq:Evar-G0}) can be applied to describe collinear magnetic structures with an arbitrary large magnetic unit cell. However, the increasing numbers of the variational parameters and, especially, the constraints [Eq.~(\ref{eq:constraints}], makes evaluation and minimization more and more difficult as the size of this unit cell grows. Thus we shall concentrate on structures with two-sublattice order. The relevance of other phases is discussed in the concluding section.

The on-site components of the ground-state Green's function on the Bethe lattice with two inequivalent sublattice in the  limit $z \to \infty$ are given by \cite{Economou}
\begin{eqnarray}
	G_\alpha(i,i;\omega) &=& 2\,\frac{ \bar \epsilon_{i\alpha}\bar \epsilon_{i+1\alpha} \pm \sqrt{ \bar \epsilon_{i\alpha}\bar \epsilon_{i+1\alpha} ( \bar \epsilon_{i\alpha}\bar \epsilon_{i+1\alpha} - W^2)  } } {W^2 q_{i\alpha}^2 \bar \epsilon_{i\alpha} }, \nonumber \\ \label{eq:def:Gll}
\end{eqnarray}
where $W= 2 t^{*}$, $\bar \epsilon_{i\alpha} = (\omega + \lambda_{i\alpha})/q_{i\alpha}^2$. In the case $\lambda_{i\alpha} =0$ and $q_{i\alpha} =1$ one simply recovers the semielliptic density of states from the jump of the Green's function at the branch cut,
\begin{equation}
	D_0(|\omega| \leq W) =  \frac{2}{\pi W^2} \sqrt{W^2 - \omega^2} .
\end{equation}
It is useful to measure energies from the ``chemical potentials'' $\mu_\alpha \equiv -(\lambda_{i\alpha}+\lambda_{i+1\alpha})/2$, and introduce the energy difference between A and B sites, $h_\alpha \equiv |\lambda_{i\alpha}-\lambda_{i+1\alpha}|/2$, and the renormalized bandwidth, $\bar W_\alpha =  W q_{i\alpha} q_{i+1\alpha}$. In terms of these parameters, the two subbands which are given by the branch cuts of the Green's function in Eq.~(\ref{eq:def:Gll}) lie in the intervals
$ \mu_\alpha-\sqrt{h_\alpha^2 + \bar W^2_\alpha} \leq \omega \leq \mu_\alpha-h_\alpha$ and $\mu_\alpha + h_\alpha \leq \omega \leq \mu_\alpha+\sqrt{h_\alpha^2 + \bar W^2_\alpha}$.

Using the Green's function in Eq.~(\ref{eq:def:Gll}) it is possible to express the local densities as $\langle \Psi_0 | \hat n_{i\alpha} | \Psi_0 \rangle = \rho_\alpha + (-1)^i m_\alpha/2$. We can get relatively simple expressions for the homogeneous and staggered parts, namely
\begin{eqnarray}
	\rho_\alpha  &\equiv& \frac{n_{i\alpha}^0 + n_{i+1 \alpha}^0}{2} = \int\limits_{-W}^{\epsilon_{F \alpha}}\!\!d\epsilon\; D_0(\epsilon) \nonumber \\
	&=& \frac{1}{2} + \frac{1}{\pi} \; \frac{\epsilon_{F \alpha}}{W} \; \sqrt{1 - \frac{|\epsilon_{F \alpha}|^2}{W^2} }  + \frac{1}{\pi} \arcsin \frac{\epsilon_{F \alpha}}{W}	\label{eq:rhoalpha_2subl}
\end{eqnarray}
and
\begin{eqnarray}
	m_\alpha &\equiv& n_{i\alpha}^0 - n_{i+1 \alpha}^0 =  \int\limits_{-W}^{-|\epsilon_{F \alpha}|}\!\!d\epsilon\; D_0(\epsilon) \frac{2\Delta_\alpha}{\sqrt{\epsilon^2+ \Delta_\alpha^2}} \nonumber \\
	&=& \frac{4}{\pi} \frac{\Delta_\alpha}{W} \sqrt{1 + \frac{\Delta_\alpha^2}{W^2} } \left( F(e_\alpha, k_\alpha) - E(e_\alpha, k_\alpha ) \right), \label{eq:malpha_2subl}
\end{eqnarray}
where $e_\alpha = \arccos(|\frac{\epsilon_{F \alpha}}{W}|)$, $k_\alpha = 1 / \sqrt{1 + \Delta_\alpha^2/W^2 }$, and the functions $F$ and $E$ are incomplete elliptic integrals of the first and second kind (defined according to Abramowitz and Stegun \cite{AbramowitzStegun}), respectively, while the roles of the parameters $\epsilon_{F \alpha}$ and $\Delta_\alpha$ are discussed below. 

The variational energy for two-sublattice long-range order can finally be written as
\begin{equation}
	E_v = \sum_\alpha \left( q_{A\alpha} q_{B\alpha} K^0_\alpha + U \frac{d_{A \alpha} + d_{B \alpha}}{2}  \right) , \label{eq:Evar}
\end{equation}
where the mean-field kinetic energy can also be expressed in a compact form,
\begin{eqnarray}
        	K^0_\alpha &=& \int\limits_{-W}^{-|\epsilon_{F \alpha}|}\!\!d\epsilon\; D_0(\epsilon) \frac{\epsilon |\epsilon|}{\sqrt{\epsilon^2+ \Delta_\alpha^2}} \nonumber \\
		&=& \phantom{+} \frac{4 W}{3 \pi} \frac{\Delta_\alpha^2}{W^2} \sqrt{1 + \frac{\Delta_\alpha^2}{W^2} } \; F(e_\alpha,k_\alpha) \nonumber \\
		&& - \frac{2 W}{3 \pi } (1 + 2 \frac{\Delta_\alpha^2}{W^2}) \sqrt{1 + \frac{\Delta_\alpha^2}{W^2} } \; E(e_\alpha,k_\alpha) \nonumber \\
		&& + \frac{2 W }{3 \pi } \sqrt{1 -  \frac{\epsilon_{F \alpha}^2}{W^2} } \; \frac{|\epsilon_{F \alpha}|}{W} \sqrt{ \frac{\epsilon_{F \alpha}^2}{W^2} + \frac{\Delta_\alpha^2}{W^2} } .
  \label{eq:K0alpha_2subl}
\end{eqnarray}
Equations (\ref{eq:rhoalpha_2subl}), (\ref{eq:malpha_2subl}), and (\ref{eq:K0alpha_2subl}) generalize the expressions corresponding to the two-component antiferromagnetic ansatz in Ref.~\cite{MetznerVollhardt} to the three-component case on the Bethe lattice.

The variational energy for a given $\rho$ and $U$ and a given set of variational parameters,  $\rho_\alpha$ (with a constraint $\sum_\alpha \rho_\alpha = 3 \rho$), $\Delta_\alpha$, $d_{i \alpha}$ and $t_i$ $ (i=A,B)$, can be calculated as follows: First, one has to solve Eq.~(\ref{eq:rhoalpha_2subl}) for each $\alpha =1,\;2,\;3$ for $\epsilon_{F \alpha}$. Then calculate $m_\alpha$ and $K^0_\alpha$ from $\Delta_\alpha$ using Eqs.~(\ref{eq:malpha_2subl}) and (\ref{eq:K0alpha_2subl}), respectively. Last, the occupations $n^0_{i \alpha} = \rho_\alpha + (-1)^i m_\alpha/2 $, $d_{i\alpha}$, and $t_i$ determine the renormalization factors $q_{i \alpha}$. In principle, one could use the $m_\alpha$ instead of the $\Delta_\alpha$ as variational parameters, but then one has to invert Eq.~(\ref{eq:malpha_2subl}). Finally, the energy has to be minimized taking also the constraints into account. Since the minimization of Eq.~(\ref{eq:Evar}) with respect to the variational parameters cannot be performed analytically, in the next section we discuss the results of the numerical optimization.

\section{Results}

\begin{figure}[ht]
\begin{center}
	\includegraphics[width=0.45\textwidth,clip=true]{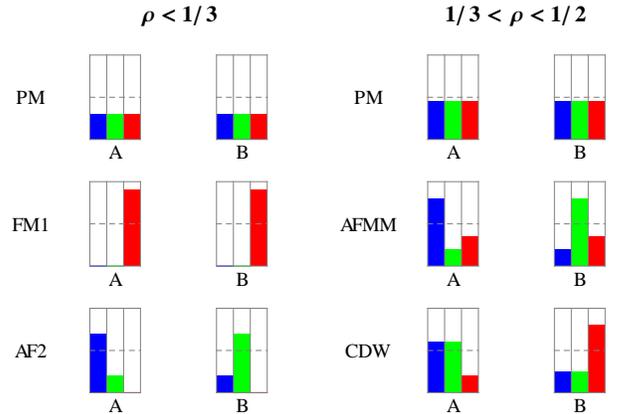}
	\caption{ \CO Examples of the sublattice occupations of the phases discussed in the text. The different colors (blue, green, red) represent the different components ($\alpha =1,\;2,\;3$, from left to right), the heights are proportional to the occupations. } \label{fig:phases}
\end{center}
\end{figure}

To explore the phase diagram in the general case with two-sublattice symmetry, we added a quadratic pe\-nal\-ty function  $\sim (\sum_{I, i\in {A,B}} \Theta(-p_{iI})p_{iI})^2$ to the variational energy and minimized it using 
different stochastic methods which gave consistent results. We found five phases which we discuss in the subsections below. The sketches of the occupations $n^0_{i\alpha}$ of the corresponding phases on the two sublattices are shown in Fig.~\ref{fig:phases}. Note that the minima are often found at the boundary of the variational space defined by the constraints.

\begin{figure}[ht]
	\centering
	\includegraphics[width=0.48\textwidth,clip=true]{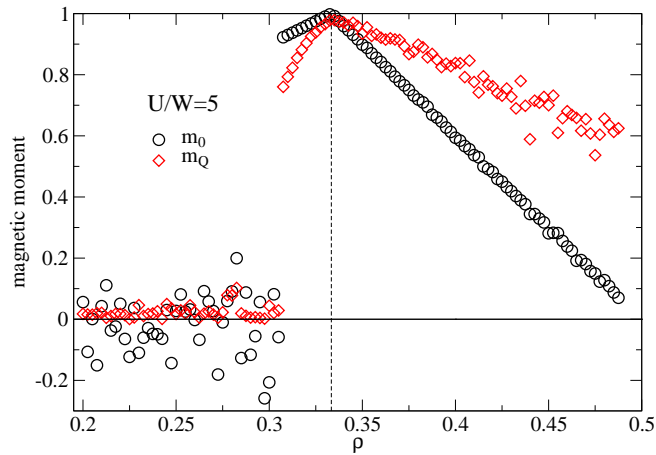}
	\caption{\label{fig:MC} \CO Raw output of parameters $m_0= \rho_1 + \rho_2 - 2 \rho_3$ and $m_Q = m_1 = -m_2$ after a MC search for the minimum for different values of $\rho$ for $U/W = 5$. For each value of $\rho$ we used $2^{30}$ random points in variational space. We see three regions with distinct behavior, corresponding to the PM, AF2, and AFMM phases.} 
\end{figure}

To check whether the correct global (rather than local) minima have been found, we also performed a brute-force random Monte Carlo search (MC) in a restricted variational space where we simply threw away configurations which violated the constraints. We assumed that $\rho_1 = \rho_2 = \rho + m_0/6$, $\rho_3 = \rho - m_0/3$, $m_1 = -m_2 = m_Q$, $m_3= 0$, $d_{A1} = d_{B2} \equiv d_1, d_{A2} = d_{B1} \equiv d_2, d_{A3}= d_{B3} \equiv d_3$ and $t_A = t_B \equiv t$, where $A$ and $B$ refer to the two sublattices. Despite the significant reduction of the variational space, four of the five phases can be described using this parametrization. In Fig.~\ref{fig:MC} a typical result of such a search is shown. 

For a precise calculation of the location of the phase transitions we performed in a third step variational calculations for each phase separately using the appropriate subset of variational parameters.

In the following we will describe each of the five relevant phases which are also sketched in Fig.~\ref{fig:phases}.

\subsection{Paramagnetic state (PM)}

At low enough values of $\rho$ or $U$, no ordering is expected and the ground state has to be a (correlated) paramagnet, defined by $m_0 = m_\alpha = 0$, $d_{i\alpha} = d,\, t_i=t$. Such a paramagnetic Gutzwiller ansatz can also be extended to the SU($N$) case with $N$ components, giving in Gutzwiller approximation
\begin{equation}
	E_v^{\rm PM} = Q^{\rm PM} N \bar \epsilon + U  {N \choose  2} d . \label{eq:def:EvarPM}
\end{equation}
Here $N \bar \epsilon$ is the total kinetic energy of the non-interacting Fermi sea with filling $\rho$.
In a simple, low density approximation (by neglecting high-order occupancies, $n_{|I|}=0$ for $|I| > 2$), we find a relatively simple expression for the renormalization factor
\begin{eqnarray}
	Q^{\rm PM} &\approx& \frac{ \rho - (N-1) d }{ \rho(1-\rho) } \times \nonumber \\
	&&  \left[ \sqrt{1 - N \rho +{{N}\choose{2}} d} + (N-1) \sqrt{d} \right]^2\!\!. \label{eq:def:QPM}
\end{eqnarray}
Note that this expression, which neglects triple and higher occupations, is exact for $N=2$. It also gives a good upper bound for the ground-state energy for low densities, $\rho < 2/N$, becoming even better as $U$ increases. The advantage of using this approximation is that we can obtain analytic expressions for general $N$. 

In particular, we see that at the commensurate filling $\rho = 1/N$ the variational energy becomes a \emph{quadratic} function of $d$, $E_v\sim d (d-d_0)$ with $d\ge 0$. The sign change of $d_0\sim U_{\rm BR}-U$ 
signals the Mott transition (found by Brinkmann and Rice \cite{SU2BrinkmanRice} for $N=2$). 
The Mott phase appears for 
\begin{equation}
	U > U_{\rm BR} = 2 \left[ 1 + \sqrt{ \frac{N}{2(N-1)} } \right]^2 N \bar \epsilon .
\end{equation}
For $N=2$, this reduces to the well-known result\cite{SU2BrinkmanRice,Fazekas}, while for $N=3$ it gives $U_{\rm BR}(N=3) \approx 3.975 W$, which matches (up to a relative error of $10^{-3}$) the numerically exact value obtained from Eq.~(\ref{eq:def:EvarPM}) when one restores the triple occupancy $t$ as a variational parameter, using
\begin{eqnarray}
	Q^{\rm PM}(N=3) &=& \left[ \sqrt{(1-3 \rho + 3 d -t)(\rho -2d+t)} \right. \nonumber \\
	&& \left. + 2\sqrt{(\rho-2d+t)(d-t)}\right. \nonumber \\
	&& \left. + \sqrt{(d-t) t} \right]^2/[\rho(1-\rho)] . \label{eq:def:QPM3}
\end{eqnarray}
Particle-hole symmetry implies that there is another Mott phase at $\rho = 1 - 1/3= 2/3$. 

\begin{figure}
	\centering
	\includegraphics[width=0.47\textwidth,clip=true]{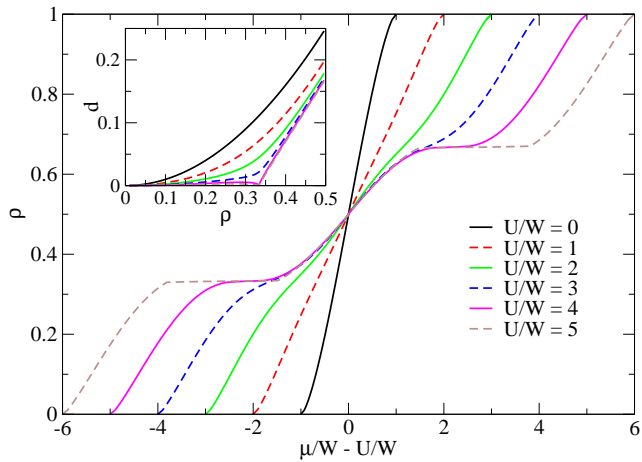}
	\caption{\label{fig:Mott} \CO The filling $\rho$ as a function of the chemical potential $\mu = \partial E^{\rm PM}/3\partial \rho$ in the paramagnetic Gutzwiller calculation, for $U/W=0,\;1,\;2,\;3,\;4,\;5$. We see clear Mott plateaus at $\rho=1/3$ and $\rho=2/3$ for $U/W \geq 3.975$. Inset: the double occupancy $d$ as a function of the filling $\rho$.  }
\end{figure}

The results of the numerical minimization of Eq.~(\ref{eq:def:EvarPM}) with Eq.~(\ref{eq:def:QPM3}) with respect to $d$ and $t$ are displayed in Fig.~\ref{fig:Mott} which shows the filling as a function of the chemical potential. The Mott insulator found for $\rho=1/3$ (and for $\rho = 2/3$) is characterized by a vanishing compressibility, $\partial \rho/\partial \mu=0$. It is instructive to compare these results to the DMFT results of Gorelik and Bl\"umer~\cite{GorelikBlumer}
who found a Mott insulator for $U \gtrsim 2.75 W$. The quantitative discrepancy to our $T=0$ result can, however, be traced back to the finite temperature, $T/W=1/40$, used in the DMFT calculation. Using the  Kot\-li\-ar - Ruck\-en\-stein slave-boson mean field theory \cite{RKslaveboson}, which is a natural generalization of the Gutzwiller approximation to finite temperatures, for the $N=3$ paramagnetic case we obtain (not shown) at $T/W=1/40$ a critical $U\approx 2.83 W$, close to the DMFT value. Despite this good agreement, one should, however, keep in mind that DMFT and the Gutzwiller calculation provide very different scenarios of how precisely the Mott transition occurs \cite{DMFT}. In reality, however, the $T=0$ Mott transition is masked by various ordered phases discussed below.

\subsection{Completely polarized ferromagnet (FM1)}

\begin{figure}
	\centering
	\includegraphics[width=0.47\textwidth,clip=true]{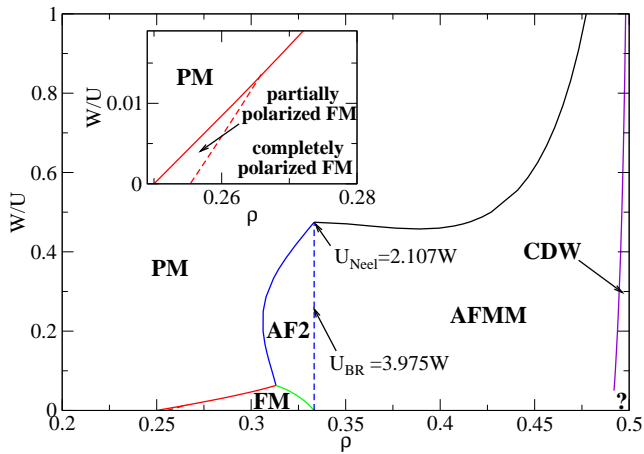}
	\caption{\label{fig:phasediag} \CO Stability regions of {\em homogeneous} phases with two-sublattice symmetry based on the Gutzwiller calculation without taking into account that phase separation can occur (see Fig.~\ref{fig:phasediagfinal} for comparison).  The inset shows the tiny region to the left of the dashed line
where one obtains a partially polarized ferromagnet with a polarization close to (but not exactly) $100$\%. All solid lines are 
first-order transitions.}
\end{figure}

For the two-component Hubbard model, it has been proven by Nagaoka in Ref.~\cite{Nagaoka} that for a class of lattices the exact ground state for $U/W \to \infty$ and a half-filled system with a single hole is a completely polarized ferromagnet. The physical argument is that all spin configurations are exactly degenerate in the  exactly half-filled system for $U=\infty$ and the kinetic energy is fully quenched. Therefore one has to ask the question which spin configuration allows for a maximal gain of kinetic energy for an additional particle or hole. For $N=2$ and $\rho=1/2$ the maximal kinetic energy $-W$ is gained only for a ferromagnetic arrangement of spins. For $N=3$ and hole doping of a $\rho=1/3$ state, that is, for the removal of a particle, the same physical argument applies but it is not valid when an extra particle is added instead. For example, a `red' particle added to a `blue-green' antiferromagnetic configuration also gains $-W$ in kinetic energy. In this case, higher order terms of order $W^2/U$ which favor antiferromagnetism will suppress ferromagnetism.

Accordingly, we find within our Gutzwiller calculation that for very large $U$ a fully polarized ferromagnetic state exists for fillings $1/4 \lesssim \rho<1/3$ (see Fig.~\ref{fig:phasediag}). Only in a small window of parameters, for example, $0.25\le \rho \le 0.255$ for $U/W=\infty$, (see inset of Fig.~\ref{fig:phasediag}) we find a partially polarized ferromagnetic state with a very large value of the polarization which jumps to zero at the first-order transition to the paramagnetic state. Note that this result is sensitive to details of the density of state.

The discussion can be generalized to arbitrary $N$ assuming a first-order transition from the paramagnet to the fully polarized ferromagnet. The energy of the ferromagnetic state, FM1, is simply given by 
\begin{equation}
E^{\rm FM1}(\rho) = \!\!\int\limits_{-W}^{\epsilon^{\rm FM1}_F}\!\!\!\!d\epsilon\;D_0(\epsilon) \epsilon, \label{eq:EFM1}
\end{equation}
which has to be compared to the paramagnetic energy discussed above. For $U=\infty$, when there are only singly occupied or empty sites, the energy of the paramagnetic state is obtained by setting $d=t=0$ in Eqns.~(\ref{eq:def:EvarPM}) and (\ref{eq:def:QPM}),
\begin{equation}
E^{\rm PM}(\rho) = \frac{1-N \rho}{1-\rho} N \!\!\int\limits_{-W}^{\epsilon^{\rm PM}_F}\!\!\!\!d\epsilon\;D_0(\epsilon) \epsilon. \label{eq:EPMi}
\end{equation}
Both $E^{\rm FM1}$ and $E^{\rm PM}$ vanish for $\rho=1/N$ when the paramagnetic solution describes a Mott insulator and the ferromagnetic one a band insulator. Furthermore, for any symmetric density of states, the two energies also coincide for $\rho=1/(N+1)$. In this case the filling of the ferromagnetic band is $N/(N+1)=1-1/(N+1)$  and thus the integrals in Eqs.~(\ref{eq:EFM1}) and (\ref{eq:EPMi}), where the band filling is $1/(N+1)$, coincide. Also the renormalization factor in Eq.~(\ref{eq:EPMi}) is $1/N$ and cancels with the factor $N$ arising from the $N$ bands.
Therefore one finds that within the Gutzwiller approximation of the SU($N$) Hubbard model the fully polarized ferromagnetic state
has a lower energy than the paramagnet for
\begin{equation}\label{stable}
\frac{1}{N+1} \leq \rho \leq \frac{1}{N}
\end{equation}
in the limit of $U\to \infty$.  

To check for the possibility of a partially polarized state, we also calculate for $U \to \infty$ the energy difference, $ E_{\rm flip}$, when a single particle has changed its color compared to the fully polarized state,
\begin{eqnarray}
E_{\rm flip}= -(1-\rho^{\rm FM}) W-\epsilon_F^{\rm FM}-\frac{1}{1-\rho^{\rm FM}} \!\int\limits_{-W}^{\epsilon^{\rm FM}_F}\!\!\!\!d\epsilon\;D_0(\epsilon) \epsilon,
\nonumber \\
\end{eqnarray}
with $\rho^{\rm FM}=\int_{-W}^{\epsilon^{\rm FM}_F}d\epsilon \, D_0(\epsilon)=N \rho$. Here
the first two terms describe the change of kinetic energy of the flipped particle and the third term reflects that the kinetic energy of all the other particles gets renormalized by the color flip. For $\epsilon_F^{\rm FM}/W>0.4317$, $E_{\rm flip}$ is positive, implying that the completely polarized state is stable. For $N=3$ and $\rho<0.255$, however, spin flips are energetically favored, as discussed above. For $N\ge 4$ on the other hand, the completely polarized state remains stable in the whole interval described by Eq.~(\ref{stable}) for $U/W \to \infty$.

\subsection{Two-component color antiferromagnet metal and insulator (AF2)}

For $\rho \leq 1/3$ filling and moderately strong $U$, the lowest energy state according to the numerical minimization can be characterized by $m_1 = -m_2 \neq 0,\,m_3=0,\,m_0 = 3\rho$, $d_3\neq 0$, implying $ \rho_3= 0$ and $d_{ 1} = d_{ 2} = t = 0$ (see Fig.~\ref{fig:phasediag}). This describes (see Fig.~\ref{fig:phases}) an antiferromagnetic metal where only two of the three components occur. With these parameters, formulas are analogous to the SU(2) antiferromagnetic state discussed in Refs.~\cite{MetznerVollhardt,SU2HubbardGutzwiller}, with a different density of states, however. At $\rho = 1/3$ one obtains a two-component antiferromagnetic insulator, which has a lower energy than the three-component PM at $\rho = 1/3$ if $U > U_{\rm Neel} \approx 2.107 \,W$. Below we will discuss that the two-component antiferromagnetic metal is never realized as a lower energy is obtained by phase separation into the antiferromagnetic Mott insulator and either a paramagnetic state 
or ferromagnetic state with lower density.

\subsection{Three component color antiferromagnetic metal (AFMM)}

For $1/3 \leq \rho \leq 1/2$ and for sufficiently strong $U$, the dominant state has parameters $m_1 = -m_2 \neq 0,\,m_3=0,\,m_0 = 3-6\rho$. Thus the first two components have exactly commensurate filling, $\rho_1 = \rho_2 = 1/2$, forming an antiferromagnetic insulator, while the third component remains metallic. This phase could thus be called an ``orbital selective'' Mott insulator. At $\rho= 1/2$, DMFT results are also available for this phase from the work of Miyatake, Inaba, and Suga~\cite{su3half}. For $U= 2.5 W, \rho=1/2$, the energy and the order parameter $m_1$ of the two methods agree within a few percent. The AFMM state was also found by Honerkamp and Hofstetter~\cite{HonerkampHofstetter} close
to $\rho=1/2$ on a square lattice using fRG.

Both AF2 and AFMM phases can be visualized as being simultaneously ferromagnetic and antiferromagnetic. Here ``ferromagnetic'' order means that two components are equally populated ($\rho_1=\rho_2$ in our parametrization) while the third one is different, breaking SU(3) symmetry. The remaining SU(2) symmetry within the first two components is broken by their antiferromagnetic ordering. Note, however, that this type of SU(3) symmetry breaking should not be confused with ferrimagnetism, that is, simultaneous ferromagnetic and antiferromagnetic order in the SU(2) case. For SU(2) symmetry, the direction of ferro- and antiferromagnetic order are not independent as the ferromagnet already breaks the SU(2) symmetry to U(1). In the resulting ``spin-flop'' phase ferro- and antiferromagnetic order are oriented perpendicular to each other as has been pointed out in the cold atom context in Ref.~\cite{gottwaldDongen}. In the SU(3) case, the larger symmetry group implies that the staggered order can still point in an arbitrary direction within the two-component subspace.

\subsection{Color density wave (CDW)}

While the previous phases can be obtained by both of the numerical methods we used, the more general minimization routine found a state with parameters $m_0 = 3 \rho - 3/2$, $\Delta_1 = \Delta_2 \geq 0$ and $\Delta_3 \leq 0$. This implies that the third component is pinned to half filling and occupies dominantly a different sublattice than the first two components, leading to a staggered modulation of the total density (see Fig.~\ref{fig:phases}). Due to the doubling of the unit cell, the third component is in a band-insulating state. As a homogeneous phase, we find that a metallic CDW phase is obtained as a minimum in a tiny doping regime close to $\rho=1/2$ (see Fig.~\ref{fig:phasediag}). Yet we will show below that this regime is unstable with respect to phase separation and only an insulating CDW state with $\rho=1/2$ is realized.

At half filling, $\rho=1/2$, the color density wave state has been found previously both by fRG on a square lattice~\cite{HonerkampHofstetter}
and by DMFT~\cite{su3half}. The energy and staggared moments we find at $\rho = 1/2, U = 2.5 W$ in the CDW phase are consistent with the DMFT calculation~\cite{su3half} within a few percent. 

Note that away from half filling, the color density wave, that is, a staggered order with $m_3 \neq m_1=m_2$, would also imply a uniform polarization of the system,
$\rho_3\neq \rho_1=\rho_2$. This should be compared to an SU(2) antiferromagnet, where staggered order does not induce ferromagnetic order. Technically, this difference between the SU(3) and SU(2) cases arises because there is no symmetry transformation which maps the relevant Gell-Mann matrix $\lambda_8$ to $-\lambda_8$ (the two matrices have a different spectrum). 

While the fRG for the $D=2$ square lattice~\cite{HonerkampHofstetter} seems to support that at half filling, the CDW is the ground state, in the DMFT study~\cite{su3half}  it is claimed that in the SU(3) case these states are degenerate. Within the Gutz\-wil\-ler calculation we found that at $\rho =1/2$, the density wave state has a slightly lower energy than the AFMM state, favoring the results from the fRG. However, for strong $U$ (even away from half filling) the whole variational energy surface $E_v = U d_{\rm tot}(1 + O(W/U))$ becomes flatter and flatter. Since different states can have the same interaction energy $\propto d_{\rm tot} = (1/2) \sum_{\alpha} (d_{A\alpha} + d_{B\alpha})$, the actual order parameters are determined by the subleading kinetic energy term. Thus the energy difference between the CDW and AFMM states decreases, and it is very difficult to track the phase transition line numerically for $U/W \to \infty$. On general grounds, however, it is clear that the AFMM can never be the ground state at half filling as the Fermi surface of the gapless third component shows perfect nesting for a particle-hole symmetric model implying that the state has to be unstable.

\subsection{Phase diagram}

In Fig.~\ref{fig:phasediag} we show which of the states has the lowest energy. As all relevant phase transitions are of first order, it is, however, clear that the phase diagram has to be modified to take phase separation into account. Close to a first-order transition one can lower the energy by allowing for heterogeneous phases, that is, by mixing phases with different densities. While in electronic systems macroscopic phase separation is not possible due to the presence of long-range Coulomb interactions, it will occur for cold atom realizations of the SU($N$) Hubbard model. The true phase diagram is obtained by using the well-known Maxwell construction for first-order transitions. For $N=2$ the relevant first-order transitions and coexistence regions have been discussed in Ref.~\cite{SU2HubbardGutzwiller}.

\begin{figure}
	\centering
	\includegraphics[width=0.47\textwidth,clip=true]{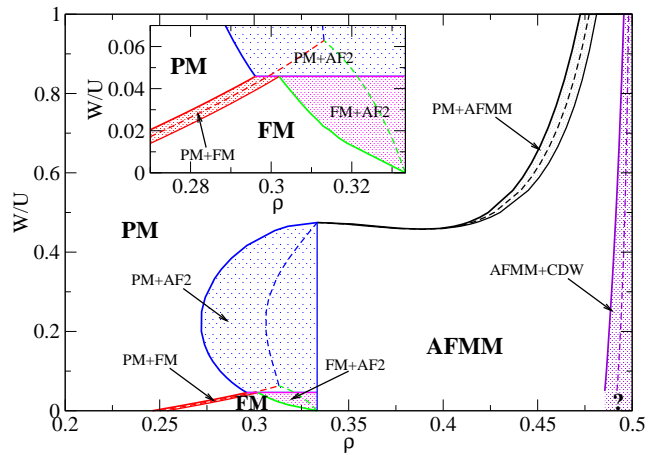}
	\caption{\label{fig:phasediagfinal} \CO Phase diagram based on the Gutzwiller calculation. All phases are separated from each other by first-order transitions and corresponding coexistence regions (shaded). The inset zooms into a region of the phase diagram where several coexistence regions meet leading to sudden jumps in the volume fractions of the corresponding phases as a function of $U$. The dashed lines represent the stability regions of the homogeneous phases shown in Fig.~\ref{fig:phasediag}, but have no direct physical significance.}
\end{figure}

In Fig.~\ref{fig:phasediagfinal} we show the phase diagram after phase separation is taken into account. Comparing this to Fig.~\ref{fig:phasediag} one realizes that both the metallic two component antiferromagnet (AF2) and the metallic color density wave (CDW) are completely wiped out by phase separation. In these regions, the commensurate AF2 insulator (and the insulating CDW) coexist with other metallic phases. A complex interplay of various phase separated regions occurs close to the ferromagnetic transition (see inset of Fig.~\ref{fig:phasediagfinal}). Note that when two different coexistence regions meet, the volume fractions of the phases jump suddenly (assuming global thermal equilibrium) when, for example, $U$ is changed. In contrast, when entering a phase separated region starting from a uniform phase, the volume fractions change always smoothly.

\section{Conclusions}

In this work we studied the ground state of the SU(3) Hubbard model using a Gutzwiller calculation. We found five phases compatible with two-sub\-lat\-ti\-ce symmetry breaking on the Bethe lattice. The resulting rather complex phase diagram is characterized by first-order transitions and various coexistence regimes. We also compared our results to two independent DMFT studies and found reasonable quantitative agreement. While we have studied the Bethe lattice in the limit of large coordination number, $z \to \infty$, we expect that the topology of the phase diagram will be very similar on a cubic lattice in three dimensions.

It is an interesting question to discuss how the phase diagram would manifest itself in a cold atom experiment assuming that low enough temperatures can be achieved to realize all phases. At least two aspects have to be considered: First, a parabolic trapping potential holds the atomic cloud together in typical experiments. Second, due to the underlying SU(3) symmetry, the number of atoms of each component is conserved and one can therefore not obtain a state with a finite net polarization starting from an unpolarized state. This is especially relevant as not only the ferromagnetic phase but also the antiferromagnetic phases AF2 and AFMM are characterized by a finite net polarization. This implies~\cite{Rapp,CherngRefael,ferro} that domains have to form such that the total net polarization of the system vanishes. The direct detection of such domains is possible by taking color-selective phase-contrast images of the cloud as long as the domain sizes are larger than the spatial resolution \cite{ferro}. 
Furthermore, staggered order can, for example, be detected by measuring noise correlations in a time-of-flight experiment \cite{noisecorrelations} or more directly by Bragg scattering \cite{Bragg}. While such approaches can distinguish, for example, two-sublattice from three-sublattice order directly, multiple domains make it very difficult to separate a color density wave from a two-sublattice antiferromagnet by measurements of correlation functions. One can, however, use the fact that the color density wave is the only phase with two-sublattice structure but no net polarization to distinguish it from the other phases.

\begin{figure}
	\centering
	\includegraphics[width=0.47\textwidth,clip=true]{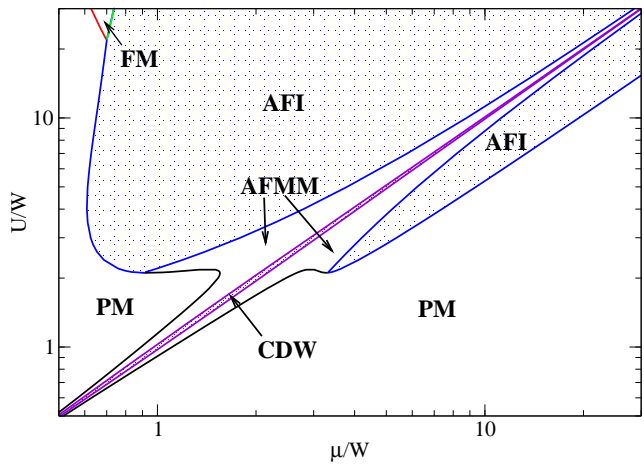}
	\caption{\label{fig:phasediagmu} \CO Phase diagram as a function of the chemical potential $\mu = \partial E/3\partial \rho $. (Note the logarithmic scaling.) While the Mott insulator region is rather robust as the Mott gap is large $\sim U$, the gap of the CDW insulator is small $O(W)$. All transitions are first order with the exception of the transition from AFI to AFMM. The AFI phase can host different magnetic structures, see the text.}
\end{figure}

To understand the properties of the system in a trap it is useful to redraw the phase diagram as a function of the chemical potential, $\mu = \partial E/3\partial \rho$, instead of $\rho$ (as shown in Fig.~\ref{fig:phasediagmu}). For a large number of atoms and a smooth confining potential $V(r)$ one can locally approximate the inhomogeneous system by a homogeneous one with a local chemical potential $\mu-V(r)$ (the so-called `local density approximation'). From the phase diagram shown in Fig.~\ref{fig:phasediagmu} one can therefore directly read off the sequence of phases expected in the presence of a trapping potential. 

For large $U$, the most dominant phase is obviously the antiferromagnetic insulator. As it is characterized by a large charge gap of order $U$ it is stable for a large range of (local) chemical potentials. The other insulating phase, the color density wave obtained for $\rho=1/2$ has, in contrast, only a small charge gap which never gets larger than $\sim\!\!0.2\,W$. The difference of the two insulators can be understood by considering the limit of vanishing hopping. While for $\rho=1/3$ the motion of all particles is locked, this is not the case for $\rho=1/2$. Therefore one expects that for $\rho=1/2$ any gap will be of order $W$ rather than $U$.

The phase diagram presented here is certainly not complete. Especially in the AFI phase at $\rho=1/3$, we expect that phases we did not discuss appear. For large $U/W$, the insulating phase at low energies is described by the SU(3) Heisenberg model.
 It has been shown~\cite{SUNHeisenberg} that the classical ground state is highly degenerate and also within the
Gutzwiller approximation a large number of ordering patterns have the same energy to leading order in $W/U$. Quantum  fluctuations (not described by the Gutzwiller wave function) can stabilize certain ordered states, and it has indeed been shown by T\'oth {\it et al.}~\cite{su3Heis} that they favor a three-sublattice ordering relative to the two-sublattice antiferromagnetic order discussed by us. Higher order terms in $W/U$ may lead to a more complex magnetic phase diagram within the Mott insulating AFI phase,
but we expect that this will not change qualitatively the interplay of the AFI phase with the other phases at finite doping. For example, a simple argument suggests that particle doping favors two-sublattice ordering (the AFMM phase) relative to three-sublattice ordering for large $U$: When a single `blue' particle is added to a `red-green' two-sublattice antiferromagnet, it can gain the full kinetic energy $-W$. This is, in contrast, not possible when a particle is added to a three-sublattice antiferromagnet where all three colors take part in the magnetic order. It is also unlikely that subtle quantum fluctuations can remove the strong first-order transition from the AFI to the PM phase obtained from our variational study.

Overall, the physics of the SU(3) Hubbard model turns out to be very different from the well-established SU(2) version in many aspects.
(i) While in the spin-1/2 Hubbard model the Mott insulating state occurs at half filling, where the paramagnetic Fermi surface is perfectly nested, the Mott insulator of the SU(3) model at $\rho=1/3$ shows no nesting in the paramagnetic state. 
(ii) Due to its larger symmetry, the Mott insulating phase can support a larger set of competing phases, for example, those with three-sublattice structure discussed above. (iii) The SU(3) Mott insulator reacts very differently to particle doping. Adding more particles does not destroy magnetic order as efficiently as in the SU(2) case: two-sublattice antiferromagnetism is not frustrated by doping as the dynamics of a third species does, to leading order, not perturb a magnetic state formed by the first two colors. This leads to the stable AFMM phase where two components remain exactly at half filling, forming an orbitally selective Mott insulator. (iv) Hole doping, in contrast, shows different physics characterized by a strong first-order transition
to the paramagnetic state. (v) Also, all other transitions from the paramagnetic state to ordered states are strongly first order and associated with large jumps of the occupations (with the exception of the weakly interacting regime very close to half filling). (vi) Finally, the physics at half filling, $\rho=1/2$, is fundamentally different. While a phase with a charge gap is also obtained in the SU(3) case, the color density wave does not arise from a Mott insulator: its gap remains finite in the large $U$ limit.

For the future, it is an interesting question how more exotic phases, like the ``chiral spin liquid''~\cite{chiralspinliquid,SUNHeisenberg}, expected for larger $N$, react to doping and whether also new types of superconductivity can be realized in repulsive SU($N$) Hubbard models.

\textbf{Acknowledgments:} We thank Gergely Zar\'and and Matthias Vojta for useful discussions. Financial support is provided by the SFB 608 and SFB TR12 of the DFG.

\appendix

\section{Cavity theory}

We shall discuss how to calculate local quantities within the Gutzwiller approximation using functional integrals and the cavity method of dynamical mean field theory. As a concrete example, we shall calculate the Gutzwiller expectation value of triple occupancies in the limit $z \to \infty$. Using the definition of the triple occupancy operator, we get
\begin{eqnarray}
	t_i &=& \frac{\langle G \vert \hat n_{i1} \hat n_{i2} \hat n_{i3} \vert G \rangle}{\langle G \vert G \rangle} \nonumber \\
	    &=&	\frac{\langle \Psi_0 \vert \lambda_t^2(i)  \hat n_{i1} \hat n_{i2} \hat n_{i3} \left( \prod_{j\neq i} \sum_I y_I(j) \hat n_{j I} \right) \vert \Psi_0 \rangle }{\langle G \vert G \rangle},\nonumber \\
\end{eqnarray}
where $y_I = \sum_{I'\subseteq I} (-1)^{|I| - | I'|} \lambda_{I'}^2$. Here a creation operator $\hat c^\dagger_{i\alpha}$ is always to the left of the corresponding annihilation operator $\hat c_{i\alpha}$, and normal ordering can be performed easily. By applying Wick's theorem to the normal ordered expression, we can express both the numerator and the denominator as a functional integral,
\begin{eqnarray}
	t_i &=& \int\!\!{\cal D}\Psi^\dagger {\cal D}\Psi \; e^{\Psi^\dagger G_0^{-1} \Psi}\!\! \nonumber \\
		&& \times \frac{  \lambda_t^2(i)  \tilde n_{i1} \tilde n_{i2} \tilde n_{i3} \left( \prod_{j\neq i} \sum_I y_I(j) \tilde n_{j I} \right) }{\int\!\!{\cal D}\Psi^\dagger {\cal D}\Psi \; e^{\Psi^\dagger G_0^{-1} \Psi} \; \prod_{j} \sum_I y_I(j) \tilde n_{j I}} \;.
\end{eqnarray}
We can simply reexponentiate all terms $j\neq i$ in the product, however, the term $i$ is missing to complete the action $S_{\rm aux}$ [defined by Eq.~(\ref{eq:def:Saux})]. To this end we define coefficients $w_I$ such that
\begin{equation}
	\left( \sum_I y_I(i) \tilde n_{iI} \right) \left( \sum_{J} w_J(i) \tilde n_{iJ} \right) \equiv 1 \label{eq:yIinverse}
\end{equation}
holds. It can be shown that $w_I$ can always be expressed in terms of $\lambda_I^2$, for example, $w_\emptyset=1/\lambda_\emptyset^2$, $w_1 = (\lambda_\emptyset^2-\lambda_1^2)/\lambda_\emptyset^4$, however, we shall never use these values explicitly.

The Gutzwiller expectation value of the triple occupancy is therefore finally expressed as an expectation value in the auxiliary theory:
\begin{eqnarray}
	t_i &=& \int\!\!{\cal D}\Psi^\dagger {\cal D}\Psi \; \frac{e^{-S_{\rm aux}}}{Z_{\rm aux}}\!\!\left[ \lambda_t^2(i)  \tilde n_{i1} \tilde n_{i2} \tilde n_{i3} \left( \sum_I w_I(i) \tilde n_{iI} \right) \right]\!\!. \nonumber \\
\end{eqnarray}
Now we integrate over all Grassmann variables except for the site $i$, which we call the cavity. Formally, this gives
\begin{eqnarray}
	t_i &=& \int\!\!d\Psi^\dagger d\Psi \; \frac{e^{-S_{\rm cav}[i]}}{Z_{\rm cav}[i]}\!\!\left[ \lambda_t^2(i)  \tilde n_{1} \tilde n_{2} \tilde n_{3} \left( \sum_I w_I(i) \tilde n_{I} \right) \right],\nonumber \\
	\label{eq:ti-cavityintegral}
\end{eqnarray}
however, the calculation of the cavity action $S_{\rm cav}[i]$ is not possible in general. But in the limit of $z\to\infty$, one finds that the contributions from other sites just renormalize the quadratic terms~\cite{DMFT}, and therefore the cavity action is
\begin{equation}
	S_{\rm cav}[i] = \Psi^\dagger (-{\cal D}^{0\;-1}[i]) \Psi + \sum_I u_I(i) \tilde n_I ,
\end{equation}
where the cavity bare propagator ${\cal D}^{0}$ has to be determined self-consistently from the condition that any local quantity defined at site $i$ in the auxiliary theory has to have the same expectation value as in the cavity theory. We shall postpone the calculation of ${\cal D}^{0}$ for the moment.

The main advantage of the cavity method is that we can calculate any finite dimensional Grassmann integrals in the cavity explicitly, that is, solve the ``impurity problem'' exactly. In contrast to dynamical mean-field theory this is always possible since the action is static. The cavity partition function is simply given by
\begin{equation}
	Z_{\rm cav} = \frac{1}{{\cal D}^{0}_1 {\cal D}^{0}_2 {\cal D}^{0}_3} \sum_I A_I ,
\end{equation}
where we introduced the notation
\begin{eqnarray}
	A_\emptyset &=& \lambda_\emptyset^2 (1-{\cal D}^{0}_1)(1-{\cal D}^{0}_2)(1-{\cal D}^{0}_3), \nonumber \\
	A_1 &=& \lambda_1^2 {\cal D}^{0}_1(1-{\cal D}^{0}_2)(1-{\cal D}^{0}_3) ,\dots \nonumber \\
	A_{\bar 1} &=& \lambda_{\bar 1}^2 (1-{\cal D}^{0}_1) {\cal D}^{0}_2 {\cal D}^{0}_3 ,\dots \nonumber \\
	A_t &=& \lambda_t^2 {\cal D}^{0}_1 {\cal D}^{0}_2 {\cal D}^{0}_3 .
\end{eqnarray}
Finally, we can perform the Grassmann integrals in Eq.~(\ref{eq:ti-cavityintegral}) and find that the triple occupancy is simply
\begin{equation}
	t_i = \frac{A_t(i)}{\sum_I A_I(i)} .  \label{eq:t-AI}
\end{equation}
To derive this expression, we do not need the explicit expressions for $u_I$ and $w_I$ in terms of the Gutzwiller parameters $\lambda_I$, just the fact that both the exponential representation $e^{-\sum_I u_I \tilde n_I}$ and the ``inverse'' $ \sum_I w_I \tilde n_I$ exist.

Repeating analogous steps for the double and single occupancies lead to expressions, for example,
\begin{equation}
	d_{i1} = \frac{\langle G \vert \hat n_{i2} \hat n_{i3} \vert G \rangle}{\langle G \vert G \rangle} = \frac{A_{\bar 1}(i) + A_t(i)}{\sum_I A_I(i)} \label{eq:d1-AI}
\end{equation}
and 
\begin{equation}
	n_{i1} = \frac{\langle G \vert \hat n_{i1} \vert G \rangle}{\langle G \vert G \rangle} = \frac{A_{1}(i) + A_{\bar 2}(i) + A_{\bar 3}(i) + A_t(i)}{\sum_I A_I(i)} . \label{eq:n1-AI}
\end{equation}

Let us now fix $\lambda_\emptyset$ and $\lambda_\alpha$ such that
\begin{eqnarray}
	{\sum_I A_I} &=& 1,\nonumber \\
	A_{1} + A_{\bar 2} + A_{\bar 3} + A_t &=& {\cal D}^0_1,\nonumber \\
	A_{2} + A_{\bar 1} + A_{\bar 3} + A_t &=& {\cal D}^0_2,\nonumber \\
	A_{3} + A_{\bar 1} + A_{\bar 2} + A_t &=& {\cal D}^0_3, \label{eq:condition-2}
\end{eqnarray}
and discuss the implications. 

First, we shall determine ${\cal D}^0$ from the self-consistency relation (``DMFT equation'') for the local propagator. This means that
\begin{equation}
	G_{ii} = \langle -\Psi_i \Psi_i^\dagger \rangle_{S_{\rm aux}} \equiv  \langle -\Psi \Psi^\dagger \rangle_{S_{\rm cav}[i]} = {\cal D}[i] . \label{eq:app:sc}
\end{equation}
The left-hand side can be calculated using Dyson's equation on the lattice,
\begin{equation}
	G_{ii} = \left[ (G_0^{-1} - \Sigma)^{-1} \right]_{ii} ,
\end{equation}
where $\Sigma$ is the proper self-energy matrix in the auxiliary field theory, which becomes site diagonal in the limit $z\to\infty$~\cite{DMFT}. The right-hand side of Eq.~(\ref{eq:app:sc}) can be calculated analytically and one finds that
\begin{equation}
	{\cal D}_\alpha = \frac{{\cal D}^0_\alpha \sum_{I: \alpha \not\in I} A_I}{{(1-\cal D}^0_\alpha) \sum_{I} A_I} .
\end{equation}
If we use the conditions in Eq.~(\ref{eq:condition-2}), we simply get 
\begin{equation}
	{\cal D}_\alpha = {\cal D}^0_\alpha .
\end{equation}
But this means that the proper self energy of the cavity theory vanishes. Furthermore, in the limit $z\to\infty$ the proper self energy in the cavity is equivalent to the self energy on the lattice~\cite{DMFT}, and therefore
\begin{equation}
	{\cal D}^0_\alpha[i] = {\cal D}_\alpha[i] = [G_{ii}]_\alpha = [G^0_{ii}]_\alpha = n^0_{i\alpha} .
\end{equation}
From Eqs.~( \ref{eq:t-AI}) -- (\ref{eq:n1-AI}) we get
\begin{eqnarray}
	n_{i\alpha} &=& {\cal D}^0_\alpha[i] = n^0_{i\alpha}, \\
	d_{i 1} &=& \lambda_{\bar 1}^2(i) (1 - n^0_{i1})n^0_{i2} n^0_{i3} +  \lambda_{t}^2(i)  n^0_{i1} n^0_{i2} n^0_{i3},  \label{eq:dG} \\
	t_{i} &=& \lambda_{t}^2(i)  n^0_{i1} n^0_{i2} n^0_{i3} \label{eq:tG} .
\end{eqnarray}

We conclude that fixing the single-occupancy parameters $\lambda_\alpha$ in such a way that the physical and the bare densities coincide, implies that the proper self energy in the auxiliary field theory vanishes for $z\to\infty$.

Also, Eqns.~(\ref{eq:cond-1}), (\ref{eq:cond-2}), (\ref{eq:dG}), and (\ref{eq:tG}) can be used to replace the Gutzwiller variational parameters by the physical occupancies. The relations take the form of ``mass laws''~\cite{BunemannGebhardWeber}, for example,
\begin{eqnarray}
	\lambda_\emptyset^2(i) &=& \frac{p_{i \emptyset}}{p^0_{i \emptyset}} \nonumber \\
				&=& \frac{1 - n^0_{i1}-n^0_{i2}-n^0_{i3}+d_{i1}+d_{i2}+d_{i3} - t_i}{(1- n^0_{i1})(1-n^0_{i2})(1-n^0_{i3})} , \nonumber \\
	\lambda_1^2(i) &=& \frac{p_{i 1}}{p^0_{i 1}} \nonumber \\
			&=& \frac{ n^0_{i1}-d_{i2}-d_{i3} + t_i}{n^0_{i1}(1-n^0_{i2})(1-n^0_{i3})} , \nonumber \\
	\lambda_{\bar 1}^2(i) &=& \frac{p_{i\bar 1}}{p^0_{i\bar 1}} = \frac{ d_{i1} - t_i}{(1-n^0_{i1})n^0_{i2} n^0_{i3}} , \nonumber \\
	\lambda_{t}^2(i) &=&  \frac{ p_{i t}}{p^0_{i t}} = \frac{ t_i}{n^0_{i1} n^0_{i2} n^0_{i3}} . \label{eq:masslaws}
\end{eqnarray}

\end{document}